\documentclass[twocolumn,trackchanges]{aastex701}

\defcitealias{2012MNRAS.425.1278W}{W2012}
\defcitealias{2021MNRAS.501.2677D}{DU2021}
\defcitealias{2021MNRAS.501.4534J}{J2021}
\defcitealias{2026arXiv260215321S}{S2026}

\shorttitle{New Observations of NGC 1624-2 Reveal Its Magnetic South Pole}
\shortauthors{Seadrow et al.}
\received{March 10, 2026}
\submitjournal{ApJ}
\begin{document}

\title{New Observations of the Strongly Magnetic O-star NGC 1624-2 Reveal Its Magnetic South Pole}

\author[orcid=0009-0002-0308-2497]{S. Seadrow}
\affiliation{Department of Physics \& Astronomy, Bartol Research Institute, University of Delaware, Newark, DE, 19714, USA}
\email{sseadrow@udel.edu} 

\author[orcid=0000-0002-5633-7548]{V. Petit}
\affiliation{Department of Physics \& Astronomy, Bartol Research Institute, University of Delaware, Newark, DE, 19714, USA}
\email{vpetit@udel.edu} 

\author[orcid=0000-0003-0936-5578]{D. Bohlender}
\affiliation{Dominion Astrophysical Observatory, Herzberg Astronomy and Astrophysics Program, National Research Council of Canada \\ 5071 West Saanich Road, Victoria, BC V9E 2E7, Canada}
\email{david.bohlender@gmail.com} 

\author[orcid=0000-0003-4062-0776]{A. David-Uraz}
\affiliation{Department of Physics, Central Michigan University, Mount Pleasant, MI, 48859, USA}
\email{adu@udel.edu} 

\author[orcid=0000-0002-0506-5124]{J. MacDonald}
\affiliation{Department of Physics \& Astronomy, Bartol Research Institute, University of Delaware, Newark, DE, 19714, USA}
\email{jimmacd@udel.edu} 

\author[orcid=0000-0003-0825-3443]{J. Ma{\'i}z Apell{\'a}niz}
\affiliation{Centro de Astrobiolog{\'i}a (CAB), CSIC-INTA, Campus ESAC. C. Bajo Del Castillo S/N, E-28 692 Madrid, Spain }
\email{jmaiz@cab.inta-csic.es}

\author[orcid=0000-0003-2580-1464]{M. Oksala}
\affiliation{Department of Physics, California Lutheran University , Thousand Oaks, CA,  91360, USA}
\affiliation{LIRA, Paris Observatory, PSL University, CNRS, Sorbonne University, Universit{\'e} Paris Cit{\'e}, 5 place Jules Janssen, Meudon, France}
\email{meo@udel.edu} 

\author[orcid=0000-0003-1387-5044]{M. Shultz}
\affiliation{Independent Researcher, Kingston, Ontario K7L 3N6, Canada.}
\email{matt.shultz@gmail.com}

\author[orcid=0000-0002-1854-0131]{G. A. Wade }
\affiliation{Department of Physics, Engineering Physics \& Astronomy, Queen's University, Kingston, Ontario K7L 3N6, Canada}
\affiliation{Department of Physics \& Space Science, Royal Military College of Canada, PO Box 17000, Station Forces, Kingston, ON K7K 7B4, Canada}
\email{Gregg.Wade@rmc.ca}

\begin{abstract}

NCG 1624-2 has the strongest detected magnetic field of all known main-sequence O-type stars. 
It was originally found that its magnetospheric emission lines followed a $\sim$5 month periodicity, and the existing line-of-sight magnetic measurements were predominantly of strong positive (north) polarity. As such, the field's geometric interpretation has been a mildly tilted (with respect to the rotational axis) dipole, such that only the magnetic north pole is visible during a rotation cycle. 

However, \citet{2026arXiv260215321S} recently reported that new magnetospheric observations no longer phased with the established ephemeris and that the period had to be decreased by a few days. \citet{2026arXiv260215321S} also found that existing magnetic measurements did not rule out a period twice as long (306.56 d). This period suggests a different magnetic configuration with a larger dipolar tilt, making both magnetic poles visible over a single rotation. Because previous spectropolarimetric observations did not have sufficient phase coverage to distinguish between the geometries, both were equally viable. 

In this paper,  we present new spectropolarimetric observations obtained specifically to resolve this ambiguity. Our new magnetic measurements have a strong negative (south) polarity, confirming that the rotational period of NGC 1624-2 is indeed nearly twice as long as previously thought. Our measurements show that both poles come within a similar angle to our line of sight and likely have roughly the same local magnetic field strength (with a dipolar strength of 15-20 kG or more depending on the inclination angle).

\end{abstract}

\keywords{Massive stars (732) --- Stellar magnetic fields (1610) --- Stellar rotation (1629) --- O stars(1137) ---Spectropolarimetry(1973)}

\section{Introduction} 
\label{sec:intro}

Magnetospheric emission in massive stars comes from the interplay between their radiatively-driven stellar winds and the large-scale (usually dipolar) magnetic field that is anchored in their photosphere \citep{2002ApJ...576..413U}. The axis of this dipole is often not aligned with the rotational axis. Thus, as the star rotates, we view a different projection of this magnetosphere on the sky \citep[e.g.][]{2008MNRAS.389..559T}. 

Spectral lines that originate in the magnetosphere will vary between ``high states'' (when one of the magnetic poles is closest to the line of sight and emission is most intense) and ``low-states'' (when the magnetic equator is closest to the line of sight and emission is more subdued). Therefore, the time interval between two high states tracks the rotation period itself if the magnetic geometry is such that only one pole appears on the visible stellar disk during the rotation (for example, if the inclination of the rotation axis $i$ with respect to the line of sight is 20$^\circ$ and the obliquity $\beta$ between the magnetic and rotation axes is, say, 10$^\circ$). The variation of the emission, usually measured by time series of equivalent width (EW) measurements, is said to be \textit{single-wave} since only one high state is observed per rotation.

Alternatively, the time interval between two high-states can track only one half of the rotation period if the magnetic geometry is such that \textit{both} magnetic poles appear on the visible stellar disk during the rotation. That said, if one magnetic pole does not come as close to the line-of-sight as the other pole, then two consecutive high states would not have the same strength of emission, and we would be able to determine that the variation of the EWs is a so-called \textit{double–wave}. However, if the geometry is such that both poles come nearly as close to the line of sight, both the \textit{north} high state and the \textit{south} high state would have similar EWs (for example if $i=\beta=90^\circ$). Therefore, the variation of the emission would not differentiate between a single-wave or double-wave magnetic configuration. Fortunately, measurements of the magnetic longitudinal field from spectropolarimetric observations can readily solve this ambiguity. In a double-wave configuration, the longitudinal magnetic field at the time of high-state emission alternates between a positive (north) and negative (south) value as the polarity of the visible pole changes. 

NGC 1624-2 \citep[spectral type O7f?p;][]{2010ApJ...711L.143W} is the main sequence O-type star with the strongest known surface magnetic field. \citet[][hereafter \citetalias{2012MNRAS.425.1278W}]{2012MNRAS.425.1278W} discovered the magnetic field with 2 spectropolarimetric observations. They also complemented these spectropolarimetric observations with 64 spectroscopic observations spanning 5 years and used the variation of spectral lines, with contributions from the magnetosphere, to establish a variation cycle with a period of 157.99 days. 

\citetalias{2012MNRAS.425.1278W} found that the EW maxima of all the high-states were the same. Combined with a positive and a near zero measurement of the longitudinal magnetic field, they interpreted the high-state recurrence timescale to be the rotational period itself, inferring a magnetic configuration that produces a single-wave EW variation of magnetospheric lines. Subsequent spectropolarimetric follow-up by \citet[][hereafter \citetalias{2021MNRAS.501.2677D}]{2021MNRAS.501.2677D} and \citet[][hereafter \citetalias{2021MNRAS.501.4534J}]{2021MNRAS.501.4534J} yielded positive or near zero longitudinal magnetic field values (from photospheric absorption lines), and continued to employ the period determined by \citetalias{2012MNRAS.425.1278W} as the rotational period. 

\citet[][hereafter \citetalias{2026arXiv260215321S}]{2026arXiv260215321S}
extended the baseline for monitoring the magnetospheric spectral lines of NGC 1624-2 from $\sim$ 5 years to 11 years (much longer than the span of the existing spectropolarimetric observations) using high-resolution spectra. They found that the EW measurements of the latest observations did not phase with the 158-d period anymore, and that the high state recurrence timescale needed to be reduced to 153 days.

More importantly, they also established that the existing spectropolarimetric observations did \textit{not} in fact rule out a magnetic configuration for which both magnetic poles appear on the visible stellar disk as the star rotates. This implies that it would be possible for the rotational period to be 306 days, and for the south magnetic pole to pass as close to the line-of-sight as the north pole does. This would imply that the magnetic geometry and strength ($i$, $\beta$, and $B_p$) could be fundamentally different from that determined by \citetalias{2012MNRAS.425.1278W}. \citetalias{2026arXiv260215321S} predicted that spectropolarimetric observations, obtained at the right time, could clearly distinguish between the single-wave or double-wave magnetic configurations. 

In this paper, we present two new spectropolarimetric observations that establish unequivocally that, indeed, the south magnetic pole of NGC 1624-2 is visible, the rotation period is nearly twice as long as previously thought, and the magnetic configuration is significantly different than that reported by \citetalias{2012MNRAS.425.1278W} and adopted by various authors since.

\section{Observations} 

\label{sec:obs}

We obtained two new spectropolarimetric observations with the ESPaDOnS spectropolarimeter at the Canada-France-Hawaii Telescope in January 2025 and December 2025. We also use the archival observations described by \citetalias{2012MNRAS.425.1278W} and \citetalias{2021MNRAS.501.2677D}, for a total of 12 measurements (given the long period, observations obtained within a few nights of each other were co-added). We note that these same archival observations were also used by \citetalias{2021MNRAS.501.4534J}, although the observations were not co-added.

The new observations were obtained with the same setup as the archival observations. The observations are high resolution ($R \sim$ 65,000), echelle spectra that acquired Stokes $I$ and Stokes $V$ covering 3,700 to 10,500 \AA. All observations were acquired with ESPaDOnS mounted on the Canada-France-Hawaii Telescope, apart from one observation that was obtained with the Narval spectropolarimeter, which is a twin of ESPaDOnS, at the Bernard Lyot Telescope. As recommended by \citetalias{2026arXiv260215321S}, the new observations were scheduled to occur during what would be a south high-state when using their 306.56 d double-wave period. Table 1 provides the observation details. 

We follow the data reduction procedure of \citetalias{2021MNRAS.501.2677D}, which we briefly summarize here. Each ESPaDOnS observation was split into two exposure sequences (each polarimetric sequence requires 4 sub-exposures with different rotations of the polarization optics), to avoid the saturation limit. The two unnormalized sequences are co-added order-by-order in the same way as \citetalias{2021MNRAS.501.2677D} (we provide our \texttt{python} implementation in the Zenodo repository). The resulting co-added observations are normalized with the same interactive IDL code used by \citetalias{2021MNRAS.501.2677D} (resulting in $I/I_c$ and Stokes $V/I_c$).

%%%%OLDER FIGURE
%\begin{figure*}
%\centering
%    \includegraphics[width=0.9\textwidth]{09-stokesI_V_phased-paper-fig.pdf}
%    \caption{Intensity and Stokes $V/I_C$ profiles of He\textsc{i} $\lambda$5015 (left) and C\textsc{iv} $\lambda$5812 (right). 
%    The vertical offset corresponds to the phase computed with the double-wave ephemeris of \citetalias{2026arXiv260215321S}.
%    The new observations are shown in purple (Jan 2025) and red (Dec 2025). 
%    The grey curve in the Stokes $V$ panels show the diagnostic Null profile (Null 1).
%    The profiles have been normalized to their local intensity continuum, as described in \S\ref{sec:Bz}. The profiles have been shifted to the stellar rest frame by subtracting the radial velocity (-30 km/s). 
%    For display purposes only, we apply a small boxcar averaging to smooth profiles.}
%    \label{fig:phased_I_and_V}

%\end{figure*}

\begin{figure*}
\centering
    \includegraphics[width=0.9\textwidth]{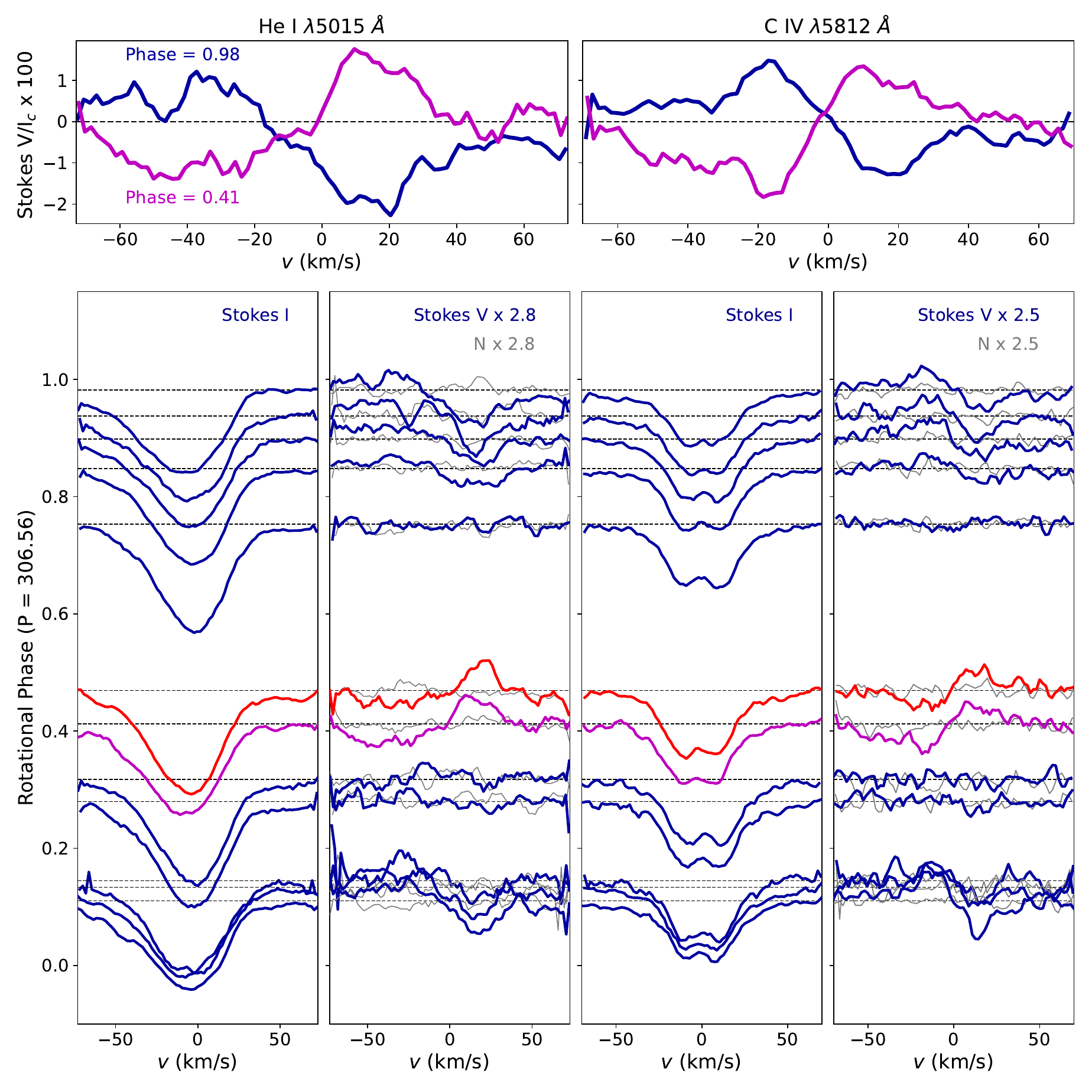}
    \caption{Intensity and Stokes $V/I_C$ profiles of He\textsc{i} $\lambda$5015 (left) and C\textsc{iv} $\lambda$5812 (right). The top panels compare Stokes $V/I_C$ (scaled by 100) from two observations at phases 0.98 (blue) and 0.41 (purple; Jan 2025), according to the double-wave ephemeris of \citetalias{2026arXiv260215321S}. The bottom panels show intensity and Stokes $V/I_C$ profiles for each line, which are offset vertically according to their phase with the same ephemeris.
    The new observations are shown in purple (Jan 2025) and red (Dec 2025). 
    The grey curves in the Stokes $V$ panels show the diagnostic Null profile.
    The profiles have been normalized to their local intensity continuum, as described in \S\ref{sec:Bz}. The profiles have been shifted to the stellar rest frame by subtracting the radial velocity ($-30$ km/s). 
    For display purposes only, we apply a small boxcar averaging to smooth profiles.}
    \label{fig:phased_I_and_V}

\end{figure*}

Figure \ref{fig:phased_I_and_V} shows the intensity and Stokes $V/I_c$ line profiles of He\textsc{i} $\lambda$5015 and C\textsc{iv} $\lambda$5812, phased with the double-wave period of \citetalias{2026arXiv260215321S}. In the top panel, we compare the Jan. 2025 observation (purple) with the archival high-state observation (blue) that yields the strongest longitudinal field measurements (for both \citetalias{2021MNRAS.501.2677D} and \citetalias{2021MNRAS.501.4534J}). The polarity of the Stokes $V$ signature for the new Jan. 2025 observation is clearly inverse to that of the previous high-state observation and of similar amplitude. 

In the bottom panel, we show all observations vertically shifted according to their phases. The new observations are shown in purple (Jan 2025) and red (Dec 2025). The polarities of the Stokes $V$ for both new observations are inverse to those of the previous high-state observations (at the top and bottom of the figure). The polarity reversal also coincides with the middle of the 306 d rotation cycle and is between low-state observations. 

These observations confirm that the double-wave rotational period proposed by \citetalias{2026arXiv260215321S} is the correct one. Furthermore, these new observations offer direct evidence for a geometry in which the magnetic south pole of NGC 1624-2 comes roughly as close to the line-of-sight as the north pole

\section{Longitudinal Magnetic Field Measurements}
\label{sec:Bz}

To characterize the magnetic field of these new observations, we measure the longitudinal magnetic field \citep[$B_\ell$, see eq 1 in ][]{2000MNRAS.313..851W} in the same individual spectral lines as \citetalias{2021MNRAS.501.2677D}: He\textsc{i} $\lambda$4471, 4713, 4921, 5015, 7281, He\textsc{ii} $\lambda$5412, C\textsc{iv} $\lambda$5801, 5812, and O\textsc{iii} $\lambda$5592.

We first normalized the spectral lines locally. We fit a linear function to two continuum regions on each side of the line profile. The regions used were not listed by \citetalias{2021MNRAS.501.2677D} – we obtained them from the authors, and we provide a list in the Zenodo archive. We also provide data files for each observation and each line profile for reproducibility. 

The calculation of $B_\ell$ requires bounds of integration; we use the same bounds as \citetalias{2021MNRAS.501.2677D} and provide them in the Zenodo repository. The calculation of $B_\ell$ also requires us to set the position in the line profile where the Doppler velocity $v$ in the $B_\ell$ equation is zero. This is, of course, straightforward to determine for a simple and symmetric line profile. However, the spectral lines of NGC 1624-2 either show some Zeeman splitting or asymmetries typical of O-type stars (often due to non-LTE effects and/or wind contamination). \citetalias{2021MNRAS.501.2677D} used the center-of-gravity (COG) of the Stokes $V$ profiles. This gives reasonable results when the Stokes $V$ signal is strong. However, this results in large variations from observation to observation, up to 30 km/s for broad and asymmetric He lines. This seems unreasonable, given that some of the computed COGs for He lines are \textit{outside} the profiles of narrower lines such as CIV. Upon closer examination, this is an issue especially for the few rotational phases where the Stokes $V$ signal is nearly zero – the COG becomes simply the center of the integration window (which is uneven relative to the core of the line because of the strong asymmetry). Given that there is no evidence of global radial velocity variation of spectral lines, we therefore opt to use a single value for the  radial velocity of all observations and spectral lines. This does not change the computed values of $B_\ell$ significantly (because the strongest Stokes V COG disparities are for phases for which $B_\ell$ is close to zero anyway), but it appears to be less \textit{ad hoc}. Comparison graphs are provided in the Zenodo repository to further demonstrate this point. Table 1 lists our recomputed $B_\ell$ for each co-added spectrum and each spectral line. 

To compute $B_\ell$, we use the \textsc{calc\_bz} function in the \textsc{specpolFlow} package \citep{2025JOSS...10.7891F}. We set the code options to \textsc{norm=1.0}, given that we have locally normalized line profiles and \textsc{cog=-30} (km/s). Finally, He\textsc{i} $\lambda$4471 is located in a region of echelle order overlap in the ESPaDOnS observations (but not for Narval). We treated both orders independently and averaged the resulting $B_\ell$, weighted by the error bar. This is also true for O\textsc{iii} $\lambda$5592 (for all observations); although \citetalias{2021MNRAS.501.2677D} only reported the measurement for the redder order, we compute the average of both orders. 

Similarly to \citetalias{2021MNRAS.501.2677D}, we take an average of the $B_\ell$ from all lines for each observation. First, we will note that we were unable to reproduce the average values reported in Table A2 of \citetalias{2021MNRAS.501.2677D} from the individual line measurements listed in their Table A3, even after obtaining the datafiles with the unrounded values. Secondly, \citetalias{2021MNRAS.501.2677D} discarded C\textsc{iv} $\lambda$5812 in their averages. Upon closer inspection, we did not find the measurements from this line to be more discrepant than those of the other lines, and we therefore include it in our weighted averages (in Table 1). Overall, these amount to only small differences and do not change our conclusions in any way (we provide a detailed documented comparison in the associated Zenodo archive).

\begin{figure*}
\centering
    \includegraphics[width=0.9\textwidth, trim={0cm 0cm 0cm 0cm}, clip]{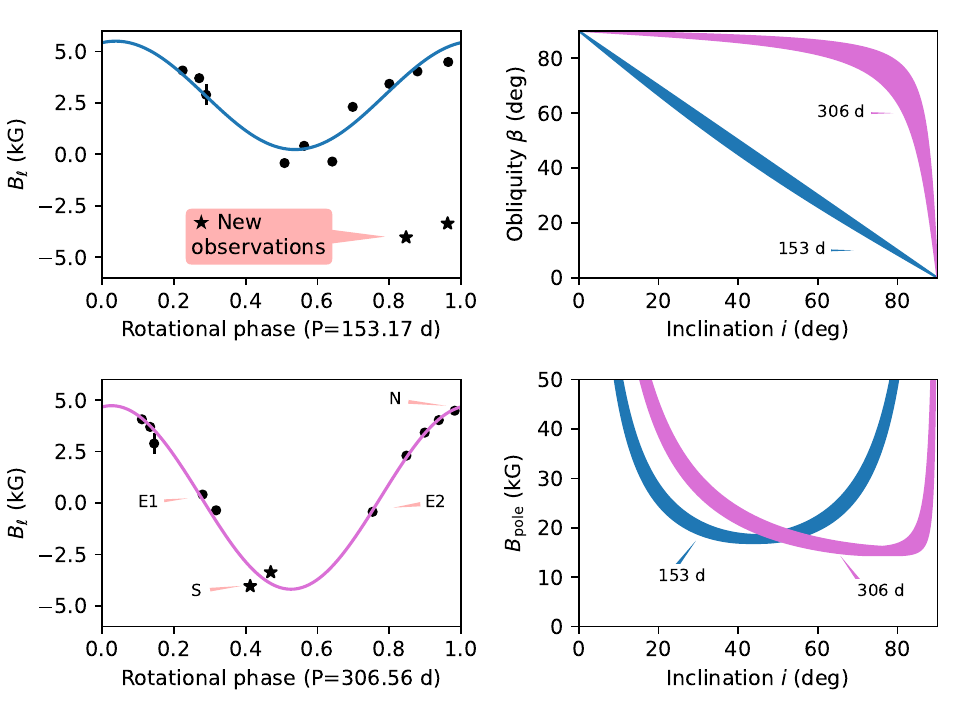}
    \caption{\label{fig:Bz}Left: Longitudinal magnetic field measurements phased with the single-wave (top) and double-wave (bottom) periods of \citetalias{2026arXiv260215321S}. In the top panel, the two new observations are indicated. In the bottom panel, the labeled observations correspond to those shown in Figure \ref{fig:CIV_split} The colored curves show a cosine model fit (that only includes the previous observations for the top panel, but all observations for the bottom panel). Right: The dipolar geometry constraints for the obliquity angle (top) and the dipolar field strength (bottom) that stem from the cosine fit as a function of the rotational inclination angle, as described in \ref{sec:Bz}. The colors match the fit curves in the left panels. 
}
\end{figure*}

The left column of Figure \ref{fig:Bz} shows the mean longitudinal field curve phased with the single-wave (top) and double-wave (bottom) magnetospheric periods of \citetalias{2026arXiv260215321S}. 

The new observations (marked with the label in the top panel) clearly show that if the single-wave magnetospheric period was correct, one would have expected a strong positive (north) longitudinal magnetic field at that rotational phase. The curve in the top left panel shows a sinusoidal fit to the measurements that excludes the new observations for reference. 
Furthermore, previous positive measurements already exist in that range of rotational phases, emphasizing that it is the period that is incorrect rather than the field being somewhat complex (this might have been reasonable to consider had these new measurements landed in one of the larger gaps in the phase coverage). 

The bottom left panel shows that the longitudinal field measurements, while phased with the double-wave magnetospheric period, have a simple sinusoidal variation, as expected from a dipolar magnetic field for which both poles come relatively close to the line of sight. 
The sinusoidal fit shown in this panel includes the new observations, and we now use them to infer some properties of the dipolar field \citep{1967ApJ...150..547P}.

The magnitudes of the minimum (south pole extremum) and maximum (north pole extremum) longitudinal magnetic field strengths are similar; the extrema of the fitted curve are $-4.2$ and $4.7$ kG, respectively. This means that the south and north magnetic poles come similarly close to our line of sight, as predicted by \citetalias{2026arXiv260215321S}, assuming a double-wave variation of the magnetospheric lines and from their fit to the previous $B_\ell$ measurements. The $r$ value \citep[$B_{\ell,\mathrm{min}}/B_{\ell,\mathrm{max}}$,][]{1967ApJ...150..547P} of $-0.89$ is therefore very close to $-1$. 

From this value, we calculate the expected obliquity angle as a function of the inclination angle \citep{1967ApJ...150..547P} and show the result in the top right panel. The shading is obtained by simply propagating an uncertainty of $\pm250 G$ (the typical error bar of our measurements) on the $B_{\ell,\mathrm{min}}$ and $B_{\ell,\mathrm{max}}$ values in our calculations. 

The implied obliquity angle is large (nearly 90$^\circ$) for most inclination angles until it becomes unconstrained at large inclination angles (i.e., a large range of $\beta$ values will give a longitudinal magnetic field variation consistent with observations when the inclination is near 90$^\circ$). For comparison, we also show the previous constraint from the $B_\ell$ variation with the shorter period (without our new observations); in that case, the magnetic geometry was such that $i+\beta\simeq45^\circ$. Therefore, we have confirmed that the magnetic geometry matches the long period option presented in \citetalias{2026arXiv260215321S}, and is thus quite different from the one adopted by \citetalias{2012MNRAS.425.1278W}, \citetalias{2021MNRAS.501.2677D}, and \citetalias{2021MNRAS.501.4534J}.

Combining the constraint between $i$ and $\beta$ with the maximum value of the longitudinal field (from the sinusoidal fit), we also calculate the inferred polar strength of the dipolar field as a function of the inclination angle \citep{1967ApJ...150..547P}, as shown in the bottom right panel of Fig. \ref{fig:Bz}.

For intermediate inclination angles, the expected $B_\mathrm{pole}$ is 15-20 kG. It increases sharply for smaller $i$ values. When $i$ approaches 90$^\circ$, the dipolar field becomes unconstrained because $\beta$ becomes unconstrained. The range of $\beta$ values that are consistent with the observations is associated with an equally large range of different dipolar field strengths; the larger values of $\beta$ require a weaker $B_\mathrm{pole}$ and a value of $r$ in the allowed range that is closest to $-1$.  We again show the previous constraint with the shorter period for comparison. Although the variation with inclination angle is slightly different with the new observations and longer period, the minimum dipolar field strength remains similar (15-20 kG).

This constraint on the dipolar field relies on the limb-darkening coefficient ($u=1-I_\mathrm{limb}/I_\mathrm{center}$). Here we show calculations for $u=0.5$ – changing the value to 0.2 (or 0.8) would shift the curve up (or down) by $\sim$2 kG. We note that using the minimum and maximum values of the data points themselves to compute $r$, instead of the peak values in the fitted model, yields the same results.

\section{Zeeman Splitting of CIV Lines }
\label{sec:modulus}

The lack of macroturbulent broadening \citep[see][]{2013MNRAS.433.2497S} and negligible rotational broadening means that Zeeman splitting is visible in some of the line profiles. The full profile variation of the photospheric C\textsc{iv} $\lambda$5812 line (for which splitting was originally reported by \citetalias{2012MNRAS.425.1278W}) is shown in Fig 1 – note that the profiles are smoothed with a small boxcar for display purposes.  

\begin{figure}
\centering
    \includegraphics[width=0.9\columnwidth]{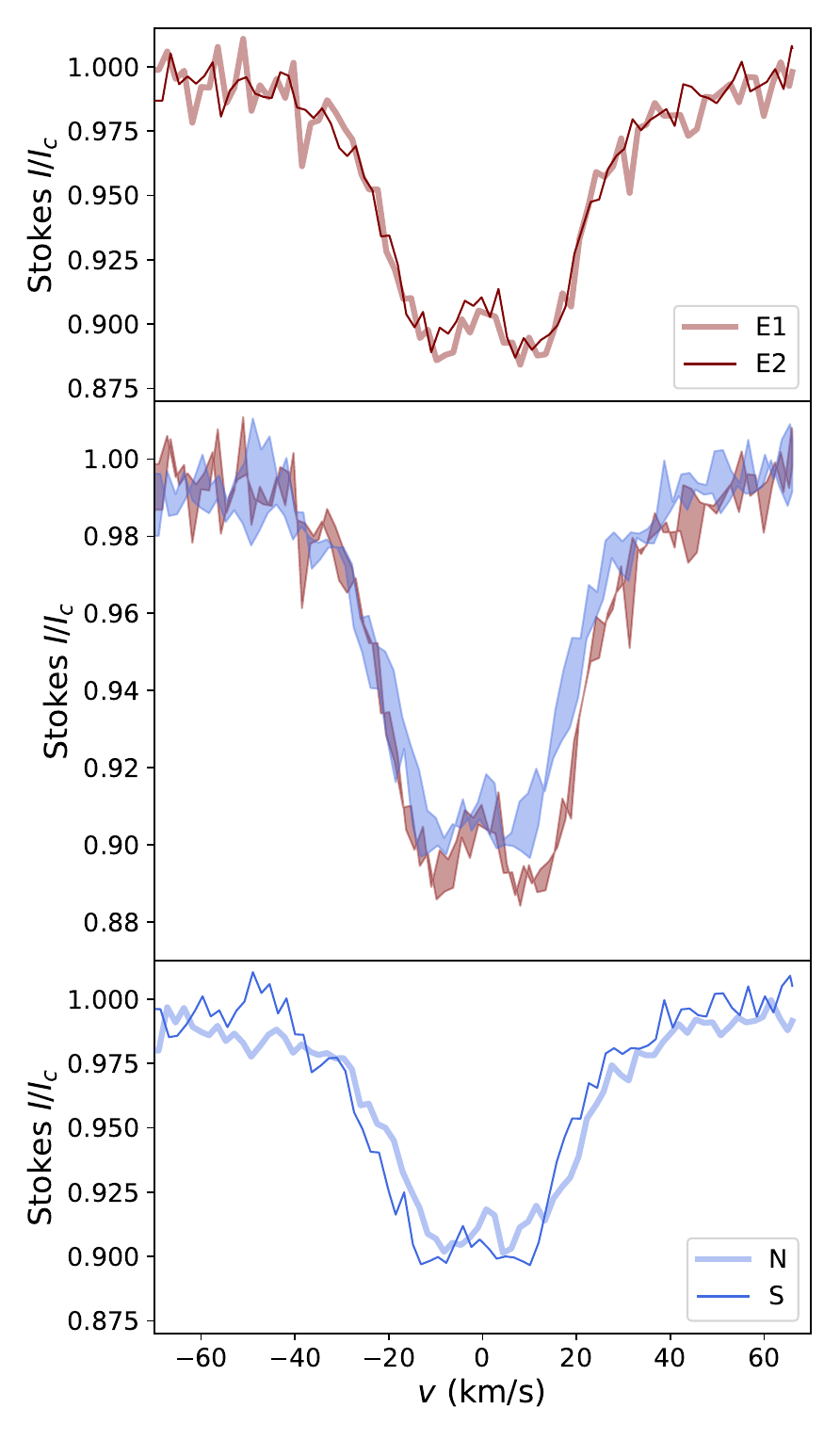}    
    \caption{\label{fig:CIV_split} The intensity profiles for C\textsc{iv} $\lambda$5812. The top panel and bottom panels compare two pole-on observations (labeled N and S in Figure \ref{fig:Bz}) and two equator-on observations (labeled E1 and E2 in Figure \ref{fig:Bz}), respectively. The middle panel shows the region between the two pole-on profiles (blue) and the region between the equator-on profiles (red). These profiles have been shifted to the stellar rest frame (by subtracting the radial velocity $-32$ km/s).}
\end{figure}

In the top panel of Figure \ref{fig:CIV_split}, we directly compare the two (non-smoothed) observations that are the closest to the two phases at which the longitudinal magnetic field curve crosses zero (labeled ``E1'' and ``E2'' in Fig. \ref{fig:Bz}), which correspond to a view of the magnetic equator in the context of a dipolar field. There are no significant differences in line depth, broadening, or in the separation of the Zeeman components.  

In the bottom panel, we compare the two observations at the extremum of the $B_\ell$ curve (labeled ``N'' and ``S'' in Fig. \ref{fig:Bz}) corresponding to the north and south pole views. We also find no significant differences between these line profiles. Therefore, the field modulus seems to be qualitatively the same, and under the assumption of a dipolar field, it implies that the strength of both poles is the same.  

We do not include direct measurements of the magnetic modulus, as it is beyond the scope of this paper. This measurement can be challenging with O-stars: in order to have accurate best-fit synthetic models for this measurement, we need to reduce the degeneracies between magnetic, thermal, and turbulent broadening. The latter can have a  significant impact on the inferred modulus.

However, as already pointed out by \citetalias{2021MNRAS.501.2677D} and \citetalias{2021MNRAS.501.4534J}, the Zeeman splitting is stronger when viewing the magnetic equator ($B_\ell \sim 0$) than when viewing the (north) magnetic pole (while $B_\ell$ is at maximum), which is not expected for a purely dipolar magnetic field. In the middle panel, we compare the envelopes of the two equatorial profiles (red) and the envelopes of the two polar profiles (blue). Both shaded regions are narrow because their respective pairs are similar, but there are noticeable differences between the two envelopes. The equatorial envelope is slightly broader, and its Zeeman components are slightly farther apart. This shows that the maximum splitting is observed for both equatorial views and that the minimum splitting is observed for both polar views.

\section{Conclusion}

\label{sec:conclusion}

We have unambiguously shown that the southern magnetic hemisphere of NGC 1624-2 is observable. Newly obtained observations show Stokes $V$ line profiles with strong negative (magnetic south) polarity. These observations yield average longitudinal field measurements of $-4.0$ and $-3.4$ kG that are on par with the positive (magnetic north) extremum of the existing measurements.

This means that the long double-wave period option presented by \citetalias{2026arXiv260215321S} (roughly twice as large as the previously adopted period), based on the variation of the magnetospheric lines, is the correct period, and there are two magnetospheric high-states and two low-states during a rotation cycle. When phased with this period, the longitudinal field varies in a single-wave sinusoid, as expected. 

This also means that the previously accepted magnetic geometry is indeed incorrect. Assuming a dipolar field, the obliquity of the magnetic axis is likely large (as inferred for most possible values of the inclination of the rotational axis), and both poles come equally close to our line of sight. Under these constraints, the minimum dipolar field is 15-20 kG and can be larger for low rotational inclinations. Taken with the rotation period, this geometry presents a real scenario in which NGC 1624-2 was spun-down due to magnetic braking (\citet{2009MNRAS.392.1022U}; \citetalias{2012MNRAS.425.1278W}) with significant obliquity instead. 

Both equation-on views have similar Zeeman splitting in the C\textsc{iv} $\lambda$5812, which points towards an azimuthal symmetry of the magnetic field. The profiles corresponding to both positive and negative extrema of the $B_\ell$ curve are also similar to each other. Their similarities lead us to reason that the magnetic modulus is roughly the same for both poles, which suggests a north-south symmetry. This also means that for both poles, the modulus is at a minimum when the longitudinal field is at an extremum, which hints at a small departure from a pure-dipolar geometry. 

In conclusion, we have drastically changed our understanding of the magnetic field geometry and rotation of the most magnetic main-sequence O-type star known to date
.

%% Please use the acknowledgment and contribution environments. This will 
%% be anonomyized when the "anonymous" style option 

\section*{Statement of Data Availability}
The ESPaDOnS observations and this papers analysis are available in ZENODO via \url{https://doi.org/10.5281/zenodo.18808408}. 

\begin{acknowledgments}

This material is based upon work supported by the National Science Foundation under Grant No. AST-2108455 (SS, VP, JM).

MO gratefully acknowledges support for this work from the National Science Foundation under Grant No. AST-2107871.

J. M. A. acknowledges support from the Spanish Government Ministerio de Ciencia e Innovación and Agencia Estatal de Investigación (10.13039/501100011033) through grant PID2022-136640~NB-C22.

GAW acknowledges support from a Natural Sciences and Engineering Research Council (NSERC) of Canada Discovery Grant.

Based on observations obtained at the Canada-France-Hawai'i Telescope (CFHT) which is operated by the National Research Council of Canada, the Institut National des Sciences de l'Univers of the Centre National de la Recherche Scientifique of France, and the University of Hawai'i. CFHT is located on Maunakea on Hawai'i Island, a mountain of considerable cultural, natural, and ecological significance. Maunakea is a sacred site to Native Hawaiians, also known as Kanaka 'Oiwi. We would like to thank the Canada-France-Hawai'i Telescope (CFHT) Operations and Software Groups for their contributions and diligence in maintaining observatory operations; the CFHT Astronomy Group for their observation coordination and data acquisition efforts; and the CFHT Finance \& Administration Group for their contributions to the management and administration of the observatory. 
\end{acknowledgments}

\bibliography{Sead_2026}{}
\bibliographystyle{aasjournalv7}

%\facilities{HST(STIS), Swift(XRT and UVOT), AAVSO, CTIO:1.3m, CTIO:1.5m, CXO}

%\software{astropy \citep{2013A&A...558A..33A,2018AJ....156..123A,2022ApJ...935..167A},  
%          Cloudy \citep{2013RMxAA..49..137F}, 
%          }


\appendix

\section{Longitudinal Magnetic Field Analysis: Single-Line and Mean Results}
\restartappendixnumbering
In this section, we list the magnetic measurements acquired from the spectropolarimetric observations.
%%%%including the bz measruements table 
%\vspace{5cm}

\thispagestyle{empty} 
\begin{sidewaystable*}[p]
    \centering
        \caption{\label{tab:BZ_results} The  $B_\ell$ (units are kG)  measurements from individual spectral lines and the weighted mean measurements, $\langle B_\ell \rangle$ measurements. We list the the observations' HJD's (minus 2450000.0 days) and phases that were computed using the revised spectroscopic ephemeris,  $ \mathrm{HJD}  =  (2455967.0 \pm  10.0) +   (306.56 \mathrm{d} \pm 1.19 \mathrm{d}) \times E $}
    \begin{tabular}{llcccccccccc}
\hline
HJD & $\phi$ & HeI 4471 & HeI 4713 & HeI 4921 & HeI 5015 & HeI 7281 & HeII 5412 & CIV 5801 & CIV 5812 & O III 5592 & $\langle B_\ell \rangle_{weighted}$ \\
\hline
5961.5152 & 0.98 & 4.6 $\pm$ 1.7 & 3.43 $\pm$ 0.54 & 5.98 $\pm$ 0.62 & 3.32 $\pm$ 0.39 & 4.73 $\pm$ 0.45 & 4.22 $\pm$ 0.56 & 6.36 $\pm$ 0.56 & 4.05 $\pm$ 0.53 & 5.53 $\pm$ 0.64 & 4.48 $\pm$ 0.18 \\
6011.3316 & 0.14 & 0.4 $\pm$ 3.2 & 2.3 $\pm$ 1.8 & 4.0 $\pm$ 2.1 & 2.3 $\pm$ 1.3 & 2.65 $\pm$ 0.98 & 5.1 $\pm$ 1.6 & 3.0 $\pm$ 1.4 & 2.8 $\pm$ 1.7 & 2.1 $\pm$ 2.0 & 2.89 $\pm$ 0.51 \\
6197.9770 & 0.75 & -0.4 $\pm$ 1.2 & -0.16 $\pm$ 0.47 & -0.64 $\pm$ 0.66 & -0.21 $\pm$ 0.35 & -0.92 $\pm$ 0.41 & -0.62 $\pm$ 0.53 & -0.78 $\pm$ 0.50 & 0.04 $\pm$ 0.48 & -0.27 $\pm$ 0.59 & -0.43 $\pm$ 0.17 \\
6533.4310 & 0.85 & 0.6 $\pm$ 1.1 & 1.47 $\pm$ 0.43 & 1.75 $\pm$ 0.55 & 2.12 $\pm$ 0.31 & 3.28 $\pm$ 0.37 & 1.96 $\pm$ 0.44 & 2.96 $\pm$ 0.40 & 2.00 $\pm$ 0.39 & 2.87 $\pm$ 0.53 & 2.30 $\pm$ 0.14 \\
6549.0645 & 0.90 & 3.3 $\pm$ 1.4 & 1.60 $\pm$ 0.51 & 2.98 $\pm$ 0.64 & 2.89 $\pm$ 0.36 & 4.04 $\pm$ 0.38 & 3.67 $\pm$ 0.52 & 4.70 $\pm$ 0.46 & 3.74 $\pm$ 0.46 & 3.18 $\pm$ 0.64 & 3.42 $\pm$ 0.16 \\
6561.0550 & 0.94 & 7.3 $\pm$ 1.8 & 2.42 $\pm$ 0.60 & 4.63 $\pm$ 0.74 & 3.46 $\pm$ 0.43 & 5.20 $\pm$ 0.43 & 4.07 $\pm$ 0.61 & 4.69 $\pm$ 0.53 & 2.94 $\pm$ 0.55 & 4.23 $\pm$ 0.72 & 4.03 $\pm$ 0.19 \\
6613.9750 & 0.11 & -0.7 $\pm$ 1.4 & 2.65 $\pm$ 0.58 & 3.59 $\pm$ 0.71 & 3.21 $\pm$ 0.41 & 5.25 $\pm$ 0.40 & 4.86 $\pm$ 0.61 & 4.13 $\pm$ 0.58 & 4.31 $\pm$ 0.54 & 5.38 $\pm$ 0.75 & 4.07 $\pm$ 0.19 \\
6621.0670 & 0.13 & 0.7 $\pm$ 1.2 & 2.00 $\pm$ 0.55 & 5.16 $\pm$ 0.71 & 2.94 $\pm$ 0.40 & 5.55 $\pm$ 0.45 & 4.02 $\pm$ 0.59 & 3.37 $\pm$ 0.52 & 3.42 $\pm$ 0.49 & 4.69 $\pm$ 0.71 & 3.70 $\pm$ 0.18 \\
6665.8525 & 0.28 & 0.6 $\pm$ 1.4 & 0.31 $\pm$ 0.59 & -0.20 $\pm$ 0.82 & 0.80 $\pm$ 0.43 & 0.89 $\pm$ 0.42 & -0.27 $\pm$ 0.63 & -0.23 $\pm$ 0.53 & 1.02 $\pm$ 0.51 & -0.63 $\pm$ 0.73 & 0.41 $\pm$ 0.19 \\
7290.5450 & 0.32 & 1.6 $\pm$ 1.6 & -1.11 $\pm$ 0.68 & -1.03 $\pm$ 0.94 & -0.51 $\pm$ 0.49 & 0.08 $\pm$ 0.41 & 1.04 $\pm$ 0.72 & -1.86 $\pm$ 0.62 & 0.41 $\pm$ 0.60 & -1.17 $\pm$ 0.83 & -0.35 $\pm$ 0.21 \\
10691.7885 & 0.41 & -6.9 $\pm$ 1.5 & -2.71 $\pm$ 0.51 & -2.89 $\pm$ 0.66 & -3.87 $\pm$ 0.40 & -4.96 $\pm$ 0.36 & -3.38 $\pm$ 0.60 & -4.08 $\pm$ 0.56 & -4.17 $\pm$ 0.57 & -4.74 $\pm$ 0.68 & -4.04 $\pm$ 0.18 \\
11015.8725 & 0.47 & -0.73 $\pm$ 0.94 & -3.21 $\pm$ 0.50 & -4.61 $\pm$ 0.68 & -2.33 $\pm$ 0.39 & -4.72 $\pm$ 0.50 & -3.18 $\pm$ 0.60 & -4.63 $\pm$ 0.56 & -2.39 $\pm$ 0.52 & -4.45 $\pm$ 0.66 & -3.36 $\pm$ 0.18 \\
\hline
\end{tabular}
\end{sidewaystable*}



\end{document}